\documentclass[12pt,a4paper]{article}
\makeatletter
\def\eqnarray{%
\stepcounter{equation}%
\let\@currentlabel=\theequation
\global\@eqnswtrue
\global\@eqcnt\z@
\tabskip\@centering
\let\\=\@eqncr
$$\halign to \displaywidth\bgroup\@eqnsel\hskip\@centering
$\displaystyle\tabskip\z@{##}$&\global\@eqcnt\@ne
\hfil$\displaystyle{{}##{}}$\hfil
&\global\@eqcnt\tw@$\displaystyle\tabskip\z@{##}$\hfil
\tabskip\@centering&\llap{##}\tabskip\z@\cr}
\makeatother
\newcommand{\fukuso}{{\mathbf C}}

\begin{document}

\title{\sl Explicit Form of the Evolution Operator of Tavis--Cummings Model 
: Three and Four Atoms Cases}
\author{
  Kazuyuki FUJII
  \thanks{E-mail address : fujii@yokohama-cu.ac.jp },\quad 
  Kyoko HIGASHIDA 
  \thanks{E-mail address : s035577d@yokohama-cu.ac.jp },\quad 
  Ryosuke KATO 
  \thanks{E-mail address : s035559g@yokohama-cu.ac.jp }\\
  Tatsuo SUZUKI
  \thanks{E-mail address : suzukita@gm.math.waseda.ac.jp },\quad 
  Yukako WADA 
  \thanks{E-mail address : s035588a@yokohama-cu.ac.jp }\\
  ${}^{*,\dagger,\ddagger,\P}$Department of Mathematical Sciences\\
  Yokohama City University, 
  Yokohama, 236--0027, 
  Japan\\
  ${}^\S$Department of Mathematical Sciences\\
  Waseda University, 
  Tokyo, 169--8555, 
  Japan\\
  }
\date{}
\maketitle
%
%
%
%
\begin{abstract}
  In this letter the explicit form of evolution operator of the 
  Tavis--Cummings model with three and four atoms is given. 
  This is an important progress in quantum optics or mathematical physics. 
\end{abstract}
%


%
%
%
%


The purpose of this paper is to give an explicit form to the evolution 
operator of Tavis--Cummings model (\cite{TC}) with some atoms. 
This model is a very important one in Quantum Optics and has been studied 
widely, see \cite{books} as general textbooks in quantum optics. 

We are studying a quantum computation and therefore want to study the model 
from this point of view, namely a quantum computation based on atoms of 
laser--cooled and trapped linearly in a cavity. We must in this model 
construct the controlled NOT gate or other controlled unitary gates to perform 
a quantum computation, see \cite{elementary-gate} as a general introduction 
to this subject. 

For that aim we need the explicit form of evolution operator of the models 
with one, two, three and four atoms (at least). As to the model of one atom or 
two atoms it is more or less known (see \cite{papers-1}), 
while as to the case of three and four atoms it has not been given as far as 
we know. Since we succeeded in finding the explicit form for three and four 
atoms cases we report it (\cite{papers-2}). 

\vspace{5mm}
The Tavis--Cummings model (with $n$--atoms) that we treat in this paper 
can be written as follows (we set $\hbar=1$ for simplicity). 
\begin{equation}
\label{eq:hamiltonian}
H=
\omega {1}_{L}\otimes a^{\dagger}a + 
\frac{\Delta}{2} \sum_{i=1}^{n}\sigma^{(3)}_{i}\otimes {\bf 1} +
g\sum_{i=1}^{n}\left(
\sigma^{(+)}_{i}\otimes a+\sigma^{(-)}_{i}\otimes a^{\dagger} \right),
\end{equation}
where $\omega$ is the frequency of radiation field, $\Delta$ the energy 
difference of two level atoms, $a$ and $a^{\dagger}$ are 
annihilation and creation operators of the field, and $g$ a coupling constant, 
and $L=2^{n}$. Here $\sigma^{(+)}_{i}$, $\sigma^{(-)}_{i}$ and 
$\sigma^{(3)}_{i}$ are given as 
\begin{equation}
\sigma^{(s)}_{i}=
1_{2}\otimes \cdots \otimes 1_{2}\otimes \sigma_{s}\otimes 1_{2}\otimes \cdots 
\otimes 1_{2}\ (i-\mbox{position})\ \in \ M(L,\fukuso)
\end{equation}
where $s$ is $+$, $-$ and $3$ respectively and 
\begin{equation}
\label{eq:sigmas}
\sigma_{+}=
\left(
  \begin{array}{cc}
    0& 1 \\
    0& 0
  \end{array}
\right), \quad 
\sigma_{-}=
\left(
  \begin{array}{cc}
    0& 0 \\
    1& 0
  \end{array}
\right), \quad 
\sigma_{3}=
\left(
  \begin{array}{cc}
    1& 0  \\
    0& -1
  \end{array}
\right), \quad 
1_{2}=
\left(
  \begin{array}{cc}
    1& 0  \\
    0& 1
  \end{array}
\right).
\end{equation}

Here let us rewrite the hamiltonian (\ref{eq:hamiltonian}). If we set 
\begin{equation}
\label{eq:large-s}
S_{+}=\sum_{i=1}^{n}\sigma^{(+)}_{i},\quad 
S_{-}=\sum_{i=1}^{n}\sigma^{(-)}_{i},\quad 
S_{3}=\frac{1}{2}\sum_{i=1}^{n}\sigma^{(3)}_{i},
\end{equation}
then (\ref{eq:hamiltonian}) can be written as 
\begin{equation}
\label{eq:hamiltonian-2}
H=
\omega {1}_{L}\otimes a^{\dagger}a + \Delta S_{3}\otimes {\bf 1} + 
g\left(S_{+}\otimes a + S_{-}\otimes a^{\dagger} \right)
\equiv H_{0}+V,
\end{equation}
which is very clear. We note that $\{S_{+},S_{-},S_{3}\}$ satisfy the 
$su(2)$--relation 
\begin{equation}
[S_{3},S_{+}]=S_{+},\quad [S_{3},S_{-}]=-S_{-},\quad [S_{+},S_{-}]=2S_{3}.
\end{equation}
However, the representation $\rho$ defined by 
$
\rho(\sigma_{+})=S_{+},\ \rho(\sigma_{-})=S_{-},\ 
\rho(\sigma_{3}/2)=S_{3}
$
is a reducible representation of $su(2)$. 

We would like to solve the Schr{\" o}dinger equation 
\begin{equation}
\label{eq:schrodinger}
i\frac{d}{dt}U=HU=\left(H_{0}+V\right)U, 
\end{equation}
where $U$ is a unitary operator (called the evolution operator). 
We can solve this equation by using the {\bf method of constant variation}. 
The result is well--known to be 
\begin{equation}
\label{eq:full-solution}
U(t)=\left(\mbox{e}^{-it\omega S_{3}}\otimes 
\mbox{e}^{-it\omega a^{\dagger}a}\right)
\mbox{e}^{-itg\left(S_{+}\otimes a + S_{-}\otimes a^{\dagger}\right)}
\end{equation}
under the resonance condition $\Delta=\omega$, 
where we have dropped the constant unitary operator for simplicity. 
Therefore 
we have only to calculate the term (\ref{eq:full-solution}) explicitly, 
which is however a very hard task \footnote{The situation is very similar to 
that of the paper quant-ph/0312060 in \cite{qudit-papers}}. 
In the following we set 
\begin{equation}
\label{eq:A}
A_{n}=S_{+}\otimes a + S_{-}\otimes a^{\dagger}
\end{equation}
for simplicity. 
We can determine\ $\mbox{e}^{-itgA}$\ for $n=1$ (one atom case), 
$n=2$ (two atoms case), $n=3$ (three atoms case) and $n=4$ (four atoms case) 
completely.

\vspace{3mm}
\par \noindent 
{\bf One Atom Case}\quad In this case $A$ in (\ref{eq:A}) is written as 
\begin{equation}
\label{eq:A-one}
A_{1}=
\left(
  \begin{array}{cc}
    0&           a \\
    a^{\dagger}& 0
  \end{array}
\right)
\equiv B_{1/2}.
\end{equation}
By making use of the simple relation 
\begin{equation}
\label{eq:relation-one}
A_{1}^{2}=
\left(
  \begin{array}{cc}
    aa^{\dagger}&   0          \\
    0           & a^{\dagger}a 
  \end{array}
\right)=
\left(
  \begin{array}{cc}
    N+1& 0  \\
    0  & N
  \end{array}
\right)
\end{equation}
with the number operator $N$ 
we have 
\begin{eqnarray}
\label{eq:solution-one}
\mbox{e}^{-itgA_{1}}
&=&
\sum_{n=0}^{\infty}\frac{(-1)^n}{(2n)!}
 \left(tg\right)^{2n}A_{1}^{2n}
-i\sum_{n=0}^{\infty}\frac{(-1)^n}{(2n+1)!}
 \left(tg\right)^{2n+1}A_{1}^{2n+1}             \nonumber \\
&=&
\left(
  \begin{array}{cc}
  \mbox{cos}\left(tg\sqrt{N+1}\right)& 
  -i\frac{\mbox{sin}\left(tg\sqrt{N+1}\right)}{\sqrt{N+1}}a  \\
  -i\frac{\mbox{sin}\left(tg\sqrt{N}\right)}{\sqrt{N}}a^{\dagger}& 
  \mbox{cos}\left(tg\sqrt{N}\right)
  \end{array}
\right).
\end{eqnarray}
We obtained the explicit form of solution. However, this form is more or less 
well--known, see for example the second book in \cite{books}.

\vspace{3mm}
\par \noindent 
{\bf Two Atoms Case}\quad In this case $A$ in (\ref{eq:A}) is written as 
\begin{equation}
\label{eq:A-two}
A_{2}=
\left(
  \begin{array}{cccc}
    0 &          a & a &           0  \\
    a^{\dagger}& 0 & 0 &           a  \\
    a^{\dagger}& 0 & 0 &           a  \\
    0 & a^{\dagger}& a^{\dagger} & 0
  \end{array}
\right).
\end{equation}

Our method is to reduce the $4\times 4$--matrix $A_{2}$ in (\ref{eq:A-two}) to 
a $3\times 3$--matrix $B_{1}$ in the following to make our calculation 
easier. 
For that aim we prepare the following matrix
\[
T=
\left(
  \begin{array}{cccc}
    0 &   1                & 0                  & 0   \\
    \frac{1}{\sqrt{2}} & 0 & \frac{1}{\sqrt{2}} & 0   \\
   -\frac{1}{\sqrt{2}} & 0 & \frac{1}{\sqrt{2}} & 0   \\
    0 &   0                &   0                  & 1
  \end{array}
\right),
\]
then it is easy to see 
\[
T^{\dagger}A_{2}T=
\left(
  \begin{array}{cccc}
    0  &                     &                     &            \\
       & 0                   & \sqrt{2}a           & 0          \\
       & \sqrt{2}a^{\dagger} & 0                   & \sqrt{2}a  \\
       & 0                   & \sqrt{2}a^{\dagger} & 0
  \end{array}
\right)\equiv 
\left(
  \begin{array}{cc}
     0 &       \\
       & B_{1} 
  \end{array}
\right)
\]
where 
$
B_{1}=J_{+}\otimes a + J_{-}\otimes a^{\dagger}
$
and $\left\{J_{+},J_{-}\right\}$ are just generators of (spin one) 
irreducible representation of (\ref{eq:sigmas}). We note that this means 
a well--known decomposition of spin 
$\frac{1}{2}\otimes \frac{1}{2}=0\oplus 1$. 

Therefore to calculate $\mbox{e}^{-itgA_{2}}$ we have only to do 
$\mbox{e}^{-itgB_{1}}$. 
Noting the relation 
\begin{eqnarray}
B_{1}^{2}&=&
\left(
  \begin{array}{ccc}
    2(N+1)             &    0     & 2a^{2}    \\
      0                & 2(2N+1)  &  0        \\
    2(a^{\dagger})^{2} &    0     & 2N
  \end{array}
\right),          \nonumber \\
\label{eq:relation-two}
B_{1}^{3}&=&
\left(
  \begin{array}{ccc}
    2(2N+3) &         &          \\
            & 2(2N+1) &          \\
            &         & 2(2N-1)
  \end{array}
\right)B_{1}\equiv DB_{1},  
\end{eqnarray}
and so 
\[
B_{1}^{2n}=D^{n-1}B_{1}^{2}\quad \mbox{for}\quad n\geq 1,\quad 
B_{1}^{2n+1}=D^{n}B_{1}\quad \mbox{for}\quad n\geq 0
\]
we obtain by making use of the Taylor expansion 
\begin{eqnarray}
\label{eq:solution-two-more(reduced)}
\mbox{e}^{-itgB_{1}}
&=&
{\bf 1}+ \sum_{n=1}^{\infty}\frac{(-1)^n}{(2n)!}
 \left(tg\right)^{2n}B_{1}^{2n}
-i\sum_{n=0}^{\infty}\frac{(-1)^n}{(2n+1)!}
 \left(tg\right)^{2n+1}B_{1}^{2n+1}        \nonumber \\
&=&
\left(
  \begin{array}{ccc}
   1+\frac{2N+2}{2N+3}f(N+1) & -ih(N+1)a & \frac{2}{2N+3}f(N+1)a^{2}  \\
   -ih(N)a^{\dagger} & 1+2f(N) & -ih(N)a                              \\
   \frac{2}{2N-1}f(N-1)(a^{\dagger})^{2} & -ih(N-1)a^{\dagger} & 
   1+\frac{2N}{2N-1}f(N-1)
  \end{array}
\right)
\end{eqnarray}
where 
\[
f(N)=\frac{-1+\mbox{cos}\left(tg\sqrt{2(2N+1)}\right)}{2},\quad 
h(N)=\frac{\mbox{sin}\left(tg\sqrt{2(2N+1)}\right)}{\sqrt{2N+1}}.
\]

\vspace{3mm}
\par \noindent 
{\bf Three Atoms Case}\quad In this case $A$ in (\ref{eq:A}) is written as 
\begin{equation}
\label{eq:A-three}
A_{3}=
\left(
  \begin{array}{cccccccc}
    0 &          a & a &           0  & a & 0 & 0 & 0          \\
    a^{\dagger}& 0 & 0 &           a  & 0 & a & 0 & 0          \\
    a^{\dagger}& 0 & 0 &           a  & 0 & 0 & a & 0          \\
    0 & a^{\dagger}& a^{\dagger} & 0  & 0 & 0 & 0 & a          \\
    a^{\dagger}& 0 & 0  &  0          & 0 & a & a & 0          \\
    0 & a^{\dagger}& 0  & 0   & a^{\dagger} &  0 & 0 & a       \\
    0 & 0 & a^{\dagger} & 0  & a^{\dagger} &  0 & 0 & a        \\
    0 & 0 & 0 & a^{\dagger} & 0 & a^{\dagger} & a^{\dagger} & 0    
  \end{array}
\right).
\end{equation}

We would like to look for the explicit form of solution like 
(\ref{eq:solution-one}) or (\ref{eq:solution-two-more(reduced)}). 
If we set 
\[
T=
\left(
  \begin{array}{cccccccc}
    0 & 0 & 0 & 0 & 1 & 0 & 0 & 0 \\
    \frac{1}{\sqrt{2}} & 0 & \frac{1}{\sqrt{6}} & 0 & 0 & 
    \frac{1}{\sqrt{3}} & 0 & 0 \\
    -\frac{1}{\sqrt{2}} & 0 & \frac{1}{\sqrt{6}} & 0 & 0 & 
    \frac{1}{\sqrt{3}} & 0 & 0 \\ 
    0 & 0 & 0 & \frac{\sqrt{2}}{\sqrt{3}} & 0 & 0 & \frac{1}{\sqrt{3}} & 0 \\
    0 & 0 & -\frac{\sqrt{2}}{\sqrt{3}} & 0 & 0 & \frac{1}{\sqrt{3}} & 0 & 0 \\
    0 & \frac{1}{\sqrt{2}} & 0 & -\frac{1}{\sqrt{6}} & 0 & 0 & 
    \frac{1}{\sqrt{3}} & 0 \\
    0 & -\frac{1}{\sqrt{2}} & 0 & -\frac{1}{\sqrt{6}} & 0 & 0 & 
    \frac{1}{\sqrt{3}} & 0 \\
    0 & 0 & 0 & 0 & 0 & 0 & 0 & 1
  \end{array}
\right),
\]
then it is not difficult to see 
\[
T^{\dagger}A_{3}T=
\left(
  \begin{array}{cccccccc}
     0 & a &   &   &   &   &   &                       \\
    a^{\dagger}& 0 &   &   &    &   &   &              \\
       &   & 0 &  a &   &   &    &                     \\
       &   & a^{\dagger} & 0 &   &   &   &             \\
       &   &   &   & 0 & \sqrt{3}a & 0 & 0             \\
       &   &   &   & \sqrt{3}a^{\dagger} & 0 & 2a & 0  \\
       &   &   &   & 0 & 2a^{\dagger} & 0 & \sqrt{3}a  \\   
       &   &   &   & 0 & 0 & \sqrt{3}a^{\dagger} & 0
  \end{array}
\right)\equiv 
\left(
  \begin{array}{ccc}
     B_{1/2} &       &        \\
           & B_{1/2} &        \\ 
           &       & B_{3/2}
  \end{array}
\right).
\]
This means a decomposition of spin $\frac{1}{2}\otimes \frac{1}{2}\otimes 
\frac{1}{2}=\frac{1}{2}\oplus \frac{1}{2}\oplus \frac{3}{2}$. 
Therefore we have only to calculate $\mbox{e}^{-itgB_{3/2}}$, which is however 
not easy. 
In this case there is no simple relation like (\ref{eq:relation-one}) or 
(\ref{eq:relation-two}), so we must find another one. 

Let us state {\bf the key lemma} for that. Noting 
\begin{eqnarray}
B_{3/2}^{2}
&=&
\left(
  \begin{array}{cccc}
    3N+3 & 0 & 2\sqrt{3}a^{2} & 0              \\
     0 & 7N+4 & 0 & 2\sqrt{3}a^{2}             \\
     2\sqrt{3}(a^{\dagger})^{2} & 0 & 7N+3 & 0 \\
     0 & 2\sqrt{3}(a^{\dagger})^{2} & 0 & 3N 
  \end{array}
\right),       \nonumber \\
B_{3/2}^{3}
&=&
\left(
  \begin{array}{cccc}
    0 & \sqrt{3}(7N+11)a & 0 & 6a^{3}                       \\
    \sqrt{3}(7N+4)a^{\dagger} & 0 & 20(N+1)a & 0            \\
    0 & 20Na^{\dagger} & 0 & \sqrt{3}(7N+3)a                \\
    6(a^{\dagger})^{3} & 0 & \sqrt{3}(7N-4)a^{\dagger} & 0
  \end{array}
\right),       \nonumber 
\end{eqnarray}
and the relations 
\[
B_{3/2}^{2n+1}=B_{3/2}B_{3/2}^{2n},\quad 
B_{3/2}^{2n+2}=B_{3/2}^{2}B_{3/2}^{2n},
\]
we can obtain $B_{3/2}^{2n}$ and $B_{3/2}^{2n+1}$ like 
\begin{eqnarray}
\label{eq:even-relation}
B_{3/2}^{2n}
&=&
\left(
  \begin{array}{cccc}
    \alpha_{n}(N+2) & 0 & 2\sqrt{3}\xi_{n}(N+2)a^{2} & 0               \\
     0 & \beta_{n}(N+1) & 0 & 2\sqrt{3}\xi_{n}(N+1)a^{2}               \\
     2\sqrt{3}\xi_{n}(N)(a^{\dagger})^{2} & 0 & \gamma_{n}(N) & 0      \\
     0 & 2\sqrt{3}\xi_{n}(N-1)(a^{\dagger})^{2} & 0 & \delta_{n}(N-1)
  \end{array}
\right),       \nonumber \\
&&  \\
\label{eq:odd-relation}
B_{3/2}^{2n+1}
&=&
\left(
  \begin{array}{cccc}
    0 & \sqrt{3}\beta_{n}(N+2)a & 0 & 6\xi_{n}(N+2)a^{3}          \\
    \sqrt{3}\beta_{n}(N+1)a^{\dagger} & 0 & 2\xi_{n+1}(N+1)a & 0  \\
    0 & 2\xi_{n+1}(N)a^{\dagger} & 0 & \sqrt{3}\gamma_{n}(N)a     \\
    6\xi_{n}(N-1)(a^{\dagger})^{3} & 0 & \sqrt{3}\gamma_{n}(N-1)a^{\dagger} & 0
  \end{array}
\right), 
\end{eqnarray}
where 
\begin{eqnarray*}
\alpha_{n}(N)&=&(v_{+}\lambda_{+}^{n}-v_{-}\lambda_{-}^{n})/(2\sqrt{d}), \quad 
\beta_{n}(N)=(w_{+}\lambda_{+}^{n}-w_{-}\lambda_{-}^{n})/(2\sqrt{d}),  \\
\gamma_{n}(N)&=&(v_{+}\lambda_{-}^{n}-v_{-}\lambda_{+}^{n})/(2\sqrt{d}), \quad 
\delta_{n}(N)=(w_{+}\lambda_{-}^{n}-w_{-}\lambda_{+}^{n})/(2\sqrt{d}), \\
\xi_{n}(N)&=&(\lambda_{+}^{n}-\lambda_{-}^{n})/(2\sqrt{d}), 
\end{eqnarray*}
and $\lambda_{\pm}\equiv \lambda_{\pm}(N)$, $v_{\pm}\equiv v_{\pm}(N)$, 
$w_{\pm}\equiv w_{\pm}(N)$, $d\equiv d(N)$ defined by 
\begin{eqnarray*}
\lambda_{\pm}(N)&=&5N\pm \sqrt{d(N)},\ 
v_{\pm}(N)=-2N-3\pm \sqrt{d(N)},\ 
w_{\pm}(N)=2N-3\pm \sqrt{d(N)},         \\
d(N)&=&16N^{2}+9.
\end{eqnarray*}

Then by making use of (\ref{eq:even-relation}) and (\ref{eq:odd-relation}) 
we have 
\begin{eqnarray}
\mbox{e}^{-itgB_{3/2}} 
&=&
 \sum_{n=0}^{\infty}\frac{(-1)^n}{(2n)!}
 \left(tg\right)^{2n}B_{3/2}^{2n}
-i\sum_{n=0}^{\infty}\frac{(-1)^n}{(2n+1)!}
 \left(tg\right)^{2n+1}B_{3/2}^{2n+1}    \nonumber \\
=&&
\left(
  \begin{array}{cccc}
    f_{2}(N+2) & -\sqrt{3}iF_{1}(N+2)a & 2\sqrt{3}h_{1}(N+2)a^{2} & 
    -6iH_{0}(N+2)a^{3}   \\
    -\sqrt{3}iF_{1}(N+1)a^{\dagger} & f_{1}(N+1) & -2iH_{1}(N+1)a & 
    2\sqrt{3}h_{1}(N+1)a^{2} \\
    2\sqrt{3}h_{1}(N)(a^{\dagger})^{2} & -2iH_{1}(N)a^{\dagger} & 
    f_{0}(N) & -\sqrt{3}iF_{0}(N)a  \\
    -6iH_{0}(N-1)(a^{\dagger})^{3} & 2\sqrt{3}h_{1}(N-1)(a^{\dagger})^{2} & 
    -\sqrt{3}iF_{0}(N-1)a^{\dagger} & f_{-1}(N-1)   
  \end{array}
\right)      \nonumber \\ 
&{}& 
\end{eqnarray}
where 
\begin{eqnarray}
f_{2}(N)&=&\left\{v_{+}(N)\mbox{cos}(tg\sqrt{\lambda_{+}(N)})-
v_{-}(N)\mbox{cos}(tg\sqrt{\lambda_{-}(N)})\right\}/(2\sqrt{d(N)}), 
\nonumber \\
f_{1}(N)&=&\left\{w_{+}(N)\mbox{cos}(tg\sqrt{\lambda_{+}(N)})-
w_{-}(N)\mbox{cos}(tg\sqrt{\lambda_{-}(N)})\right\}/(2\sqrt{d(N)}), 
\nonumber \\
f_{0}(N)&=&\left\{v_{+}(N)\mbox{cos}(tg\sqrt{\lambda_{-}(N)})-
v_{-}(N)\mbox{cos}(tg\sqrt{\lambda_{+}(N)})\right\}/(2\sqrt{d(N)}), 
\nonumber \\
f_{-1}(N)&=&\left\{w_{+}(N)\mbox{cos}(tg\sqrt{\lambda_{-}(N)})-
w_{-}(N)\mbox{cos}(tg\sqrt{\lambda_{+}(N)})\right\}/(2\sqrt{d(N)}), 
\nonumber \\
h_{1}(N)&=&\left\{\mbox{cos}(tg\sqrt{\lambda_{+}(N)})-
\mbox{cos}(tg\sqrt{\lambda_{-}(N)})\right\}/(2\sqrt{d(N)}),
\nonumber \\
F_{1}(N)&=&\left\{\frac{w_{+}(N)}{\sqrt{\lambda_{+}(N)}}
\mbox{sin}(tg\sqrt{\lambda_{+}(N)})-
\frac{w_{-}(N)}{\sqrt{\lambda_{-}(N)}}
\mbox{sin}(tg\sqrt{\lambda_{-}(N)})\right\}/(2\sqrt{d(N)}), 
\nonumber \\
F_{0}(N)&=&\left\{\frac{v_{+}(N)}{\sqrt{\lambda_{-}(N)}}
\mbox{sin}(tg\sqrt{\lambda_{-}(N)})-
\frac{v_{-}(N)}{\sqrt{\lambda_{+}(N)}}
\mbox{sin}(tg\sqrt{\lambda_{+}(N)})\right\}/(2\sqrt{d(N)}), 
\nonumber \\
H_{1}(N)&=&\left\{\sqrt{\lambda_{+}(N)}
\mbox{sin}(tg\sqrt{\lambda_{+}(N)})-
\sqrt{\lambda_{-}(N)}
\mbox{sin}(tg\sqrt{\lambda_{-}(N)})\right\}/(2\sqrt{d(N)}),
\nonumber \\
H_{0}(N)&=&\left\{\frac{1}{\sqrt{\lambda_{+}(N)}}
\mbox{sin}(tg\sqrt{\lambda_{+}(N)})-
\frac{1}{\sqrt{\lambda_{-}(N)}}
\mbox{sin}(tg\sqrt{\lambda_{-}(N)})\right\}/(2\sqrt{d(N)}).
\nonumber 
\end{eqnarray}

\vspace{3mm}
\par \noindent 
{\bf Four Atoms Case}\quad In this case $A_{4}$ in (\ref{eq:A}) is written as 
\begin{equation}
\label{eq:A-four}
A_{4}=
\left(
  \begin{array}{cccccccccccccccc}
    0 & a & a & 0 & a & 0 & 0 & 0 & a &   &   &   &   &   &   &   \\
    a^{\dagger}& 0 & 0 & a & 0 & a & 0 & 0 &   & a &   &   &   &   &   &  \\
    a^{\dagger}& 0 & 0 & a & 0 & 0 & a & 0 &   &   & a &   &   &   &   &  \\
    0 & a^{\dagger}& a^{\dagger} & 0 & 0 & 0 & 0 & a &   &   &   & a &  &  
    &   &    \\
    a^{\dagger}& 0 & 0 & 0 & 0 & a & a & 0 &   &   &   &   & a &   &  &   \\
    0 & a^{\dagger}& 0 & 0 & a^{\dagger} & 0 & 0 & a &  &  &  &  &  & a &  
    &  \\
    0 & 0 & a^{\dagger} & 0  & a^{\dagger} & 0 & 0 & a &  &  &  &  &  &  & 
    a &   \\
    0 & 0 & 0 & a^{\dagger} & 0 & a^{\dagger} & a^{\dagger} & 0 &  &  &  &  & 
    &  &  & a  \\
    a^{\dagger}&  &  &  &  &  &  &  & 0 & a & a & 0 & a & 0 & 0 & 0  \\
      & a^{\dagger}&  &  &  &  &  &  & a^{\dagger}& 0 & 0 & a & 0 & a & 0 & 0 
    \\
      &  & a^{\dagger}&  &  &  &  &  & a^{\dagger}& 0 & 0 & a & 0 & 0 & a & 0 
    \\
      &  &  & a^{\dagger}&  &  &  &  & 0 & a^{\dagger}& a^{\dagger} & 0 & 0 & 
    0 & 0 & a \\
      &  &  &  & a^{\dagger}&  &  &  & a^{\dagger}& 0 & 0 & 0 & 0 & a & a & 0 
    \\
      &  &  &  &  & a^{\dagger}&  &  & 0 & a^{\dagger}& 0 & 0 & a^{\dagger} & 
    0 & 0 & a \\
      &  &  &  &  &  & a^{\dagger}&  & 0 & 0 & a^{\dagger} & 0  & a^{\dagger} 
    & 0 & 0 & a \\
      &  &  &  &  &  &  & a^{\dagger} & 0 & 0 & 0 & a^{\dagger} & 0 & 
    a^{\dagger} & a^{\dagger} & 0 
  \end{array}
\right).
\end{equation}

If we set $T$ as 
\begin{eqnarray*}
&&T= \\
&&\hspace{-5mm}\\
&&
\left(
\begin{array}{cccccccccccccccc}
0 & 0 & 0 & 0 & 0 & 0 & 0 & 0 & 0 & 0 & 0 & 1 & 0 & 0 & 0 & 0 \\
0 & \frac{1}{\sqrt{2}} & 0 & 0 & 0 & \frac{1}{\sqrt{6}} & 0 & 0 & 
\frac{1}{2\sqrt{3}} & 0 & 0 & 0 & \frac{1}{2} & 0 & 0 & 0 \\
0 & -\frac{1}{\sqrt{2}} & 0 & 0 & 0 & \frac{1}{\sqrt{6}} & 0 & 0 & 
\frac{1}{2\sqrt{3}} & 0 & 0 & 0 & \frac{1}{2} & 0 & 0 & 0 \\
0 & 0 & 0 & 0 & \frac{1}{\sqrt{3}} & 0 & \frac{1}{\sqrt{3}} & 0 & 0 & 
\frac{1}{\sqrt{6}} & 0 & 0 & 0 & \frac{1}{\sqrt{6}} & 0 & 0 \\
0 & 0 & 0 & 0 & 0 & -\sqrt{\frac{2}{3}} & 0 & 0 & \frac{1}{2\sqrt{3}} & 0 & 
0 & 0 & \frac{1}{2} & 0 & 0 & 0 \\
\frac{1}{2} & 0 & \frac{1}{2} & 0 & -\frac{1}{2\sqrt{3}} & 0 & 
-\frac{1}{2\sqrt{3}} & 0 & 0 & \frac{1}{\sqrt{6}} & 0 & 0 & 0 & 
\frac{1}{\sqrt{6}} & 0 & 0 \\
-\frac{1}{2} & 0 & -\frac{1}{2} & 0 & -\frac{1}{2\sqrt{3}} & 
0 & -\frac{1}{2\sqrt{3}} & 0 & 0 & \frac{1}{\sqrt{6}} & 0 & 0 & 
0 & \frac{1}{\sqrt{6}} & 0 & 0 \\
0 & 0 &  0 & 0 & 0 & 0 & 0 & 0 & 0 & 0 & \frac{\sqrt{3}}{2} & 
0 & 0 & 0 & \frac{1}{2} & 0 \\
0 & 0 & 0 & 0 & 0 & 0 & 0 & 0 & -\frac{\sqrt{3}}{2} & 0 & 
0 & 0 & \frac{1}{2} & 0 & 0 & 0 \\
-\frac{1}{2} & 0 & \frac{1}{2} & 0 & -\frac{1}{2\sqrt{3}} & 0 & 
\frac{1}{2\sqrt{3}} & 0 & 
0 & -\frac{1}{\sqrt{6}} & 0 & 0 & 0 & \frac{1}{\sqrt{6}} & 0 & 0 \\
\frac{1}{2} & 0 & -\frac{1}{2} & 0 & -\frac{1}{2\sqrt{3}} & 0 & 
\frac{1}{2\sqrt{3}} & 0 & 0 & -\frac{1}{\sqrt{6}} & 0 & 0 & 
0 & \frac{1}{\sqrt{6}} & 0 & 0 \\
0 & 0 & 0 & 0 & 0 & 0 & 0 & \sqrt{\frac{2}{3}} & 0 & 0 & 
-\frac{1}{2\sqrt{3}} & 0 & 0 & 0 & \frac{1}{2} & 0 \\
0 & 0 & 0 & 0 & \frac{1}{\sqrt{3}} & 0 & -\frac{1}{\sqrt{3}} & 
0 & 0 & -\frac{1}{\sqrt{6}} & 
0 & 0 & 0 & \frac{1}{\sqrt{6}} & 0 & 0 \\
0 & 0 & 0 & \frac{1}{\sqrt{2}} & 0 & 0 & 0 & -\frac{1}{\sqrt{6}} & 
0 & 0 & -\frac{1}{2\sqrt{3}} & 
0 & 0 & 0 & \frac{1}{2} & 0 \\
0 & 0 & 0 & -\frac{1}{\sqrt{2}} & 0 & 0 & 0 & -\frac{1}{\sqrt{6}} & 
0 & 0 & -\frac{1}{2\sqrt{3}} & 0 & 0 & 0 & \frac{1}{2} & 0 \\
0 & 0 & 0 & 0 & 0 & 0 & 0 & 0 & 0 & 0 & 0 & 0 & 0 & 0 & 0 & 1
\end{array}
\right) 
\end{eqnarray*}
then it is not difficult to see 
\begin{eqnarray*}
&& T^{\dagger} A_4 T=\\
&& \hspace{-10mm}
\left(
\begin{array}{cccccccccccccccc}
	0&&&&&&&&&&&&&&& \\
	&0&{\sqrt{2}} a&0&&&&&&&&&&&& \\
	&{\sqrt{2}} a^{\dagger}&0&{\sqrt{2}} a&&&&&&&&&&&& \\
	&0&{\sqrt{2}} a^{\dagger}&0&&&&&&&&&&&& \\
	&&&&0&&&&&&&&&&& \\
	&&&&&0&{\sqrt{2}} a&0&&&&&&&& \\
	&&&&&{\sqrt{2}} a^{\dagger}&0&{\sqrt{2}} a&&&&&&&& \\
	&&&&&0&{\sqrt{2}} a^{\dagger}&0&&&&&&&& \\
	&&&&&&&&0&{\sqrt{2}} a&0&&&&& \\
	&&&&&&&&{\sqrt{2}} a^{\dagger}&0&{\sqrt{2}} a&&&&& \\
	&&&&&&&&0&{\sqrt{2}} a^{\dagger}&0&&&&& \\
	&&&&&&&&&&&0&2 a&0&0&0 \\
	&&&&&&&&&&&2 a^{\dagger}&0&{\sqrt{6}} a&0&0 \\
	&&&&&&&&&&&0&{\sqrt{6}} a^{\dagger}&0&{\sqrt{6}} a&0 \\
	&&&&&&&&&&&0&0&{\sqrt{6}} a^{\dagger}&0&2 a \\
	&&&&&&&&&&&0&0&0&2 a^{\dagger}&0 
\end{array}
\right)  \\
&&\equiv 0\oplus B_{1}\oplus 0\oplus B_{1}\oplus B_{1}\oplus B_{2}. 
\end{eqnarray*}
This means a well--known decomposition of spin 
$
\frac{1}{2}\otimes \frac{1}{2}\otimes \frac{1}{2}\otimes \frac{1}{2}
=0\oplus 1\oplus 0\oplus 1\oplus 1\oplus 2 
$.

Since we have calculated $\mbox{e}^{-itgB_{1}}$ in 
(\ref{eq:solution-two-more(reduced)}) (\cite{papers-2}) we have only to 
do $\mbox{e}^{-itgB_{2}}$, which is of course hard. 
The proof is similar to that of the three atoms case, so we state only the 
result. 
\begin{eqnarray}
&&\mbox{exp}\left(-itgB_{2} \right)= \nonumber \\
&&{} \nonumber \\
&&
\left(
\begin{array}{ccccc}
f_2(N+2) & 0 & h_1(N+2)a^2 & 0 & k_0(N+2) a^4 \\
0 & f_1(N+1) & 0 & h_0(N+1)a^2 & 0 \\
h_1(N) (a^{\dagger})^2 & 0 & f_0(N) & 0 & h_{-1}(N)a^2 \\
0 & h_0(N-1) (a^{\dagger})^2 & 0 & f_{-1}(N-1) & 0 \\
k_0(N-2) (a^{\dagger})^4 & 0 & h_{-1}(N-2) (a^{\dagger})^2 & 0 & f_{-2}(N-2)
\end{array}
\right)+     \nonumber \\
&&{}         \nonumber \\
&& \hspace{-13mm}
\left(
\begin{array}{ccccc}
 0 & -2iF_1(N+2)a & 0 & -2iH_0(N+2)a^3 & 0 \\
 -2iF_1(N+1) a^{\dagger} & 0 & -\frac{i}{2}H_1(N+1)a & 0 & -2iH_0(N+1)a^3 \\
 0 & -\frac{i}{2}H_1(N) a^{\dagger} & 0 & -\frac{i}{2}H_{-1}(N)a & 0 \\
 -2iH_0(N-1) (a^{\dagger})^3 & 0 & -\frac{i}{2}H_{-1}(N-1) a^{\dagger} & 0 & 
 -2iF_{-1}(N-1)a \\
 0 & -2iH_0(N-2) (a^{\dagger})^3 & 0 & -2iF_{-1}(N-2) a^{\dagger} & 0
\end{array}
\right)     \nonumber \\
&& 
\end{eqnarray}
where 
\begin{eqnarray*}
&&f_2(N) = 
1+4(N-1)\{ (u_+/\lambda_+) (\cos tg\sqrt{\lambda_+}-1)
 -(u_-/\lambda_-) (\cos tg\sqrt{\lambda_-}-1) \} 
 /\sqrt{d} \\
&&f_1(N) = 
(u_{+} \cos tg\sqrt{\lambda_{+}}-u_{-} \cos tg\sqrt{\lambda_{-}})
 /\sqrt{d}, \\
&&f_0(N) = 
1+2\{ (v_{+}w_{+}/\lambda_{+}) (\cos tg\sqrt{\lambda_{+}}-1)
 -(v_{-}w_{-}/\lambda_{-}) (\cos tg\sqrt{\lambda_{-}}-1) \}
 /\sqrt{d}, \\
&&f_{-1}(N) =
 (u_{+} \cos tg\sqrt{\lambda_{-}}
 -u_{-} \cos tg\sqrt{\lambda_{+}})/\sqrt{d}, \\
&&f_{-2}(N) = 
1+4(N+2) \{ (u_{+}/\lambda_{-}) (\cos tg\sqrt{\lambda_{-}}-1)
 -(u_{-}/\lambda_{+}) (\cos tg\sqrt{\lambda_{+}}-1) \} /\sqrt{d}, \\
&&h_1(N) = 
 2\{ (v_{+}/\lambda_{+}) (\cos tg\sqrt{\lambda_{+}}-1)
  -(v_{-}/\lambda_{-}) (\cos tg\sqrt{\lambda_{-}}-1) \} /\sqrt{d}, \\
&&h_0(N) =
 (\cos tg\sqrt{\lambda_{+}}-\cos tg\sqrt{\lambda_{-}})/\sqrt{d}, \\
&&h_{-1}(N) =
 2 \{ (w_{+}/\lambda_{+}) (\cos tg\sqrt{\lambda_{+}}-1)
 -(w_{-}/\lambda_{-}) (\cos tg\sqrt{\lambda_{-}}-1) \} /\sqrt{d}, \\
&&k_0(N) =
 4 \{ (1/\lambda_{+}) (\cos tg\sqrt{\lambda_{+}}-1)
 -(1/\lambda_{-}) (\cos tg\sqrt{\lambda_{-}}-1) \} /\sqrt{d}, 
\end{eqnarray*}
and
\begin{eqnarray*}
&&F_1(N) =
\{(u_{+}/\sqrt{\lambda_{+}}) \sin tg\sqrt{\lambda_{+}}-
  (u_{-}/\sqrt{\lambda_{-}}) \sin tg\sqrt{\lambda_{-}}\}/\sqrt{d}, \\
&&F_{-1}(N) =
\{(u_{+}/\sqrt{\lambda_{+}}) \sin tg\sqrt{\lambda_{-}}-
  (u_{-}/\sqrt{\lambda_{-}}) \sin tg\sqrt{\lambda_{+}}\}/\sqrt{d}, \\
&&H_1(N) =
2\{(v_{+}/\sqrt{\lambda_{+}}) \sin tg\sqrt{\lambda_{+}}-
   (v_{-}/\sqrt{\lambda_{-}}) \sin tg\sqrt{\lambda_{-}}\}/\sqrt{d}, \\
&&H_0(N) =
\{(1/\sqrt{\lambda_{+}}) \sin tg\sqrt{\lambda_{+}}-
  (1/\sqrt{\lambda_{-}}) \sin tg\sqrt{\lambda_{-}}\}/\sqrt{d}, \\
&&H_{-1}(N) =
2\{(w_{+}/\sqrt{\lambda_{+}}) \sin tg\sqrt{\lambda_{+}}-
   (w_{-}/\sqrt{\lambda_{-}}) \sin tg\sqrt{\lambda_{-}}\}/\sqrt{d}
\end{eqnarray*}
, and $d\equiv d(N)$, $\lambda_{\pm}\equiv \lambda_{\pm}(N)$, 
$u_{\pm}\equiv u_{\pm}(N)$, $v_{\pm}\equiv v_{\pm}(N)$ and 
$w_{\pm}\equiv w_{\pm}(N)$ defined by 
\begin{eqnarray*}
&&\lambda_{\pm}(N) = 10N+5 \pm 3\sqrt{d(N)},\ 
u_{\pm}(N) = \frac{1}{2}(-3 \pm \sqrt{d(N)}),                     \\
&&v_{\pm}(N) = \frac{\sqrt{3}}{\sqrt{2}}(2N-1 \pm \sqrt{d(N)}),\ 
w_{\pm}(N) = \frac{\sqrt{3}}{\sqrt{2}}(2N+3 \pm \sqrt{d(N)}),     \\
&&d(N) = 4N^2+4N+9. 
\end{eqnarray*}

\vspace{5mm}
A comment is in order. 
We note that in the process of calculation we used {\bf Mathematica} to the 
fullest (a calculation by hand might be ``painful"). 
We would like to generalize the results in this paper to the cases of 
more than four atoms. However, it is not easy to perform a calculation due to 
some severe technical reasons. 
There is a (big ?) gap between the four atoms and the five ones.

\vspace{10mm}
We obtained the explicit form of evolution operator of the Tavis--Cummings 
model for three and four atoms cases. This is a big progress in quantum optics 
or mathematical physics \footnote{To obtain an explicit solution for some 
(intersting) model is still important}. Therefore, many applications 
to quantum physics or mathematical physics will be expected, see for example 
papers in \cite{papers-1}. 

We can also apply the result(s) to a quantum computation based on atoms of 
laser--cooled and trapped linearly in a cavity \cite{FHKW}. 

We conclude this paper by making a comment. The Tavis--Cummings model 
is based on (only) two energy levels of atoms. However, an atom has in general 
infinitely many energy levels, so it is natural to use this possibility. 
We are also studying a quantum computation based on multi--level systems of 
atoms (a qudit theory) \cite{qudit-papers}. Therefore we would like to extend 
the Tavis--Cummings model based on two--levels to a model based on 
multi--levels. This is a very challenging task.



\begin{thebibliography}{99}
%
\bibitem{TC}
M. Tavis and F. W. Cummings : 
\newblock Exact Solution for an N--Molecule--Radiation--Field Hamiltonian, 
\newblock Phys. Rev. 170(1968), 379 ; 
%
E. T. Jaynes and F. W. Cummings : 
\newblock Comparison of Quantum and Semiclassical Radiation Theories with 
Applications to the Beam Maser, 
\newblock Proc. IEEE 51(1963), 89 ; 
%
R. H. Dicke : 
\newblock Coherence in Spontaneous Radiation Processes, 
\newblock Phys. Rev, 93(1954), 99. 
%
\bibitem{books}
L. Allen and J. H. Eberly : 
\newblock Optical Resonance and Two--Level Atoms, 
\newblock Wiley, New York, 1975\ ; 
%
P. Meystre and M. Sargent III : 
\newblock Elements of Quantum Optics (third edition), 
\newblock Springer--Verlag, 1990\ ; 
%
Claude Cohen--Tannoudji, J. Dupont--Roc and G. Grynberg : 
\newblock Atom--Photon Interactions ; Basic Processes and Applications, 
\newblock Wiley, New York, 1998. 
%
\bibitem{elementary-gate}
K. Fujii : 
\newblock Introduction to Grassmann Manifolds and Quantum Computation, 
\newblock J. Applied Math, 2(2002), 371, 
\newblock quant-ph/0103011\ ; 
%
A. Barenco, C. H. Bennett, R. Cleve, D. P. Vincenzo, N. 
Margolus, P. Shor, T. Sleator, J. Smolin and H. Weinfurter : %
\newblock Elementary gates for quantum computation, 
\newblock Phys. Rev. A 52(1995), 3457, 
\newblock quant-ph/9503016. 
%
\bibitem{papers-1}
M. Orszag, R. Ramirez, J. C. Retamal and C. Saavedra : 
\newblock Quantum cooperative effects in a micromaser, 
\newblock Phys. Rev. A 49 (1994), 2933\ ; 
%
M. S. Kim, J. Lee, D. Ahn and P. L. Knight : 
\newblock Entanglement induced by a single-mode heat environment, 
\newblock Phys. Rev. A 65 (2002), 040101, 
\newblock quant-ph/0109052\ ; 
%
C. Genes, P. R. Berman and a. G. Rojo : 
\newblock Spin squeezing via atom -- cavity field coupling, 
\newblock quant-ph/0306205\ ; 
%
\bibitem{papers-2}
K. Fujii, K. Higashida, R. Kato and Y. Wada : 
\newblock Explicit Form of Solution of Two Atoms Tavis-Cummings Model, 
\newblock quant-ph/0403008\ ; 
%
K. Fujii, K. Higashida, R. Kato, T. Suzuki and Y. Wada : 
\newblock Explicit Form of the Evolution Operator for the Three Atoms 
Tavis-Cummings Model, 
\newblock quant-ph/0404034\ ; 
%
K. Fujii, K. Higashida, R. Kato, T. Suzuki and Y. Wada : 
\newblock Explicit Form of the Evolution Operator for the Four Atoms 
Tavis-Cummings Model, 
\newblock quant-ph/0406184.
%
\bibitem{FHKW}K. Fujii, K. Higashida, R. Kato and Y. Wada : 
\newblock Cavity QED and Quantum Computation in the Weak Coupling Regime, 
\newblock quant-ph/0407014. 
%
\bibitem{qudit-papers}
K. Fujii : 
\newblock Exchange Gate on the Qudit Space and Fock Space, 
\newblock J. Opt. B : Quantum Semiclass. Opt, 5(2003), S613, 
\newblock quant-ph/0207002\ ; 
%
K. Fujii : 
\newblock Quantum Optical Construction of Generalized Pauli and 
Walsh--Hadamard Matrices in Three Level Systems, 
\newblock quant-ph/0309132\ ; 
%
K. Fujii, K. Higashida, R. Kato and Y. Wada : 
\newblock N Level System with RWA and Analytical Solutions Revisited, 
\newblock quant-ph/0307066\ ; 
%
K. Fujii, K. Higashida, R. Kato and Y. Wada : 
\newblock A Rabi Oscillation in Four and Five Level Systems, 
\newblock quant-ph/0312060\ ; 
%
K. Funahashi : 
\newblock Explicit Construction of Controlled--U and Unitary Transformation 
in Two--Qudit, 
\newblock quant-ph/0304078. 
%
\end{thebibliography}
\end{document}